\documentclass[preprint,prd,aps]{revtex4}

\usepackage{graphicx,dcolumn,booktabs,bm}
\usepackage{indentfirst}
\usepackage{multirow}
\usepackage{diagbox} 
\usepackage{amssymb,bm,mathrsfs,bbm,amscd}
\usepackage[tbtags]{amsmath}
\usepackage{lastpage}
\usepackage{mathrsfs}
\usepackage{xcolor}
\usepackage{verbatim}
\usepackage{setspace} 
\newcommand{\RNum}[1]{\uppercase\expandafter{\romannumeral #1\relax}} 

\begin{document}



\title{Sensitivity study of  anomalous $HZZ$ couplings at future Higgs  factory }

\author{%
       Hua-Dong Li$^{1;2}$\footnote{E-mail: lihd@ihep.ac.cn}%
 ~ Cai-Dian L\"{u}$^{1;2}$\footnote{ E-mail: lucd@ihep.ac.cn}%
  ~Lian-You Shan$^{1}$\footnote{E-mail: shanly@ihep.ac.cn}%
}

\address{%
$^1$ Institute of High Energy Physics, CAS, Beijing 100049, China \\
$^2$ School of Physics, University of Chinese Academy of Sciences, Beijing 100049, China
}

\begin{abstract}
We study the sensitivity of constraining the model independent Higgs-Z-Z coupling under effective theory up to dimension-6 operators at the future Higgs factory. Utilizing the current conceptual design parameters of the Circular Electron Positron Collider, we give the experimental limits for the model independent operators by the total Higgsstrahlung cross section and angular distribution of Z boson decay in the Higgs factory.  Especially, we give   very small  sensitivity limit for the CP violation  parameter $\tilde g$, which will be a clear window to test the Standard Model and look for New Physics signal.
\end{abstract}

\maketitle




\section{Introduction}

Since the large hadron collider (LHC) announced the discovery of a scalar-like resonance~\cite{Chatrchyan:2012xdj,Aad:2012tfa},  many subsequent measurements confirmed that it is just the Higgs particle  as the  last brick of Standard Model (SM) ~\cite{CMS:2018lkl}.  Among these measurements, the generic  Higgs coupling to vector gauge bosons 
presented the largest ($\sim 7\%$) deviation from  SM prediction, but its experimental error is so big ($ \pm 56\%$) that the good agreements with SM still stand tenable. Both the  non-explained phenomena like dark matter and the theoretical tension like hierarchy (naturalness) problem still keep the extensions to SM necessary.
 The interaction between Higgs scalar and vector  gauge bosons is a key ingredient
   for the underlying  nature of spontaneously breaking of electroweak gauge symmetry. In addition to the suggested experiments on kinematic distributions at LHC Run2  \cite{Artoisenet:2013puc,Brehmer:2015rna,Corbett:2015ksa},   future Higgs factories are  under considerations, such as  the Circular Electron Positron Collider (CEPC) in China, the International Linear Collider (ILC) in Japan, the Compact Linear Collider (CLIC) and Future Circular Collider (FCC-ee) in Europe~\cite{Fan:2014vta,Durieux:2017rsg}. Along this line, many pre-analysis have been put forward~\cite{Fujii:2017vwa,Ruan:2014xxa}, to unveil the nature of gauge boson and    Higgs couplings.
These lepton colliders will accumulate events with full kinematics and less backgrounds at high luminosity, which will support precision tests in the Higgs sector. Such it become indispensable to unfold and utilize the events ${\emph shapes}$  as the details presented by data to explore the subtleties in Higgs properties.

Theoretically, an Effective Field Theory  approach is adopted with the so called Strongly Interacting Light Higgs (SILH) scenario \cite{Giudice:2007fh,Contino:2013kra} on the HZZ couplings. This model-independent description however consists of 12 independent operators for a single HZZ vertex, it is not practical to extract so many Wilson coefficients in experiments or even on the  above mentioned lepton colliders with luminosity up to several thousands of femtobarn. So   further     compression are described in  \cite{Contino:2013kra,Alloul:2013naa} where only four phenomenological  parameters  are involved.  Many   works  have been done on theoretical analysis \cite{Hagiwara:1993sw,Beneke:2014sba,Stolarski:2012ps},  and  on the running experiment of LHC to demonstrate how to constrain these 4 parameters with  the distributions of polarization angle and azimuthal angle~\cite{Bolognesi:2012mm,Englert:2015hrx,Nakamura:2017ihk}.
Cross section sensitivity studies of anomalous Higgs couplings have   been performed at the LHC and electron-positron colliders  in \cite{Anderson:2013afp}. In ref. \cite{Khanpour:2017cfq}, the authors discussed also the angular distribution sensitivity of Z boson and Higgs decays at the electron-position collider at the energy of 350GeV and 500 GeV. Since both of the current circular electron position collider and the international linear collider are designed at the center of mass energy around 240GeV,
it is necessary to do the sensitivity study of the Higgs production in detail at the specific design of detector.
The sensitivity study of cross sections  has been done  in ref.~\cite{Chen:2016zpw}. And in ref.\cite{Craig:2015wwr}, the authors did sensitivity study on some asymmetry parameters of the angular distributions of these effective operators.

In this work, we will investigate the polarization angle of Z decay associated in Higgsstrahlung~\cite{Denner:1992bc,Kilian:1995tr}, also the azimuthal angle of the Z decay  on future leptonic colliders. Our work pays more attention to the physics/parameters to be   extracted from these angle distributions, by the sensitivity study relying on the characteristics of detector design at the future electron position collider.
Differing from most of previous studies which came mainly in theoretical fashion, it is worthy to stress that, in this work the potential systematical errors from experiment have been quoted in the Pearson $\chi^2$, which are assumed to be at the same level of the statistical one.

The paper is organized as follows: In Sec. \RNum{2}, analytical formulas of angular distribution and CP-violation terms for $Z$ decays to 2 leptons affected by new coupling $HZZ$ are presented. Sec. \RNum{3} gives our numerical limit on the sensitivity of new physics parameters on CEPC. Sec. \RNum{4} is the summary.

\section{New physics effects on HZZ coupling}

The generic effective Hamiltonian of $HZZ$ sector is written as \cite{Contino:2013kra}
\begin{eqnarray}
\mathcal{L}_{HZZ}&=&-\frac{{1}}{4}g_{1}Z_{\mu\nu}Z^{\mu\nu}h-g_{2}Z_{\nu}\partial_{\mu}Z^{\mu\nu}h+ \nonumber \\
&&g_{3}Z_{\mu}Z^{\mu}h-\frac{{1}}{4}\widetilde{g} Z_{\mu\nu}\widetilde{Z}^{\mu\nu}h ,
\label{effH}
\end{eqnarray}
from which,  the effective Feynman rule can be derived as
\begin{eqnarray}
V_{\mu\nu} &=& ig_{\mu\nu}[ g_0 + g_{3}+g_{2}(p_{3}^{2}+p_{2}^{2})+g_{1}(p_{2}\cdot p_{3})]-\nonumber \\
&&i[\frac{1}{2}g_{1}(p_{3}^{\mu}p_{2}^{\nu}+p_{2}^{\mu}p_{3}^{\nu})+g_{2}(p_{2}^{\mu}p_{2}^{\nu}+p_{3}^{\mu}p_{3}^{\nu})-\nonumber \\
&&\widetilde{g}\epsilon_{\mu\nu\rho\sigma}p_{3}^{\rho}p_{2}^{\sigma}].
\label{Feynrule}
\end{eqnarray}
In this parametrization,  $g_0 = e M_Z /(c_w s_w ) $ is the HZZ coupling in Standard model. Taking the convention of \cite{Alloul:2013naa}, $g_3$ is a small fraction defined with an implicit unit of $g_0$, while $g_1, g_2, {\tilde g} $ are also small fractions defined with an implicit unit $e^2 /( g_0 s^2_w c^4_w )$ so that the interaction are consistent in dimensions of mass.    The new type of coupling $g_1, g_2, {\tilde g} $ should  also be smaller than the SM one, since most of the experimental data are consistent with the SM up to now.
The number of free parameters in new physics is then reduced from 12 \cite{Alloul:2013naa} to only 4 while keeping a sufficiently general structure in interaction between Higgs and vector bosons. 

\begin{figure}
\begin{center}
\includegraphics[width=8cm]{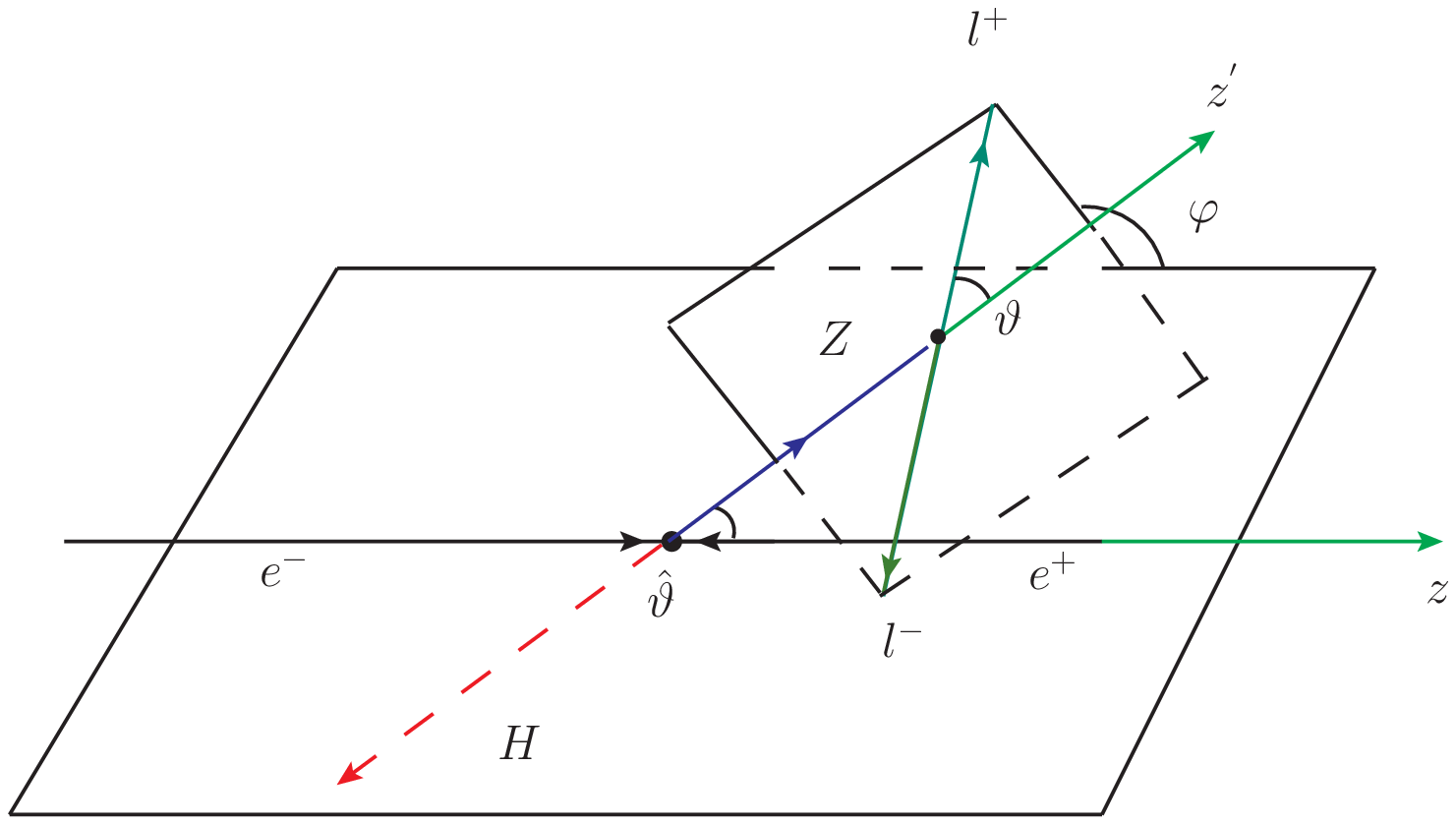}
\caption{\label{anglesdefi}Kinematics of   $e^+e^- \to H Z(l^+l^-)$. }
\end{center}
\end{figure}

In the Higgs factory of lepton collider,  the on-shell Z boson and  Higgs boson are produced simultaneously through a virtual Z boson, after electron and position annihilation.  Both the Z and Higgs boson are going to decay promptly. We will focus only on the Z decay to a pair of leptons, since they are the kinds of particles with the highest detection efficiency and carrying on the polarization message of Z boson. The kinematics of this process is illustrated in Fig. \ref{anglesdefi}. Above-mentioned new physics coupling of $HZZ$ beyond SM  will make    the   $ e^+ e^- \to Z^* \to  HZ$ cross section different from SM, which has been discussed before. Obviously the complicated new physics structure in eq.(\ref{Feynrule}) will also change the polarization fraction of the Z boson, making the angle distributions of the final lepton pairs different from the SM case.

The momenta and the helicities of incoming (anti)electron and outgoing bosons are defined through :
\begin{equation}
e^- (p_1 , \sigma_1 ) + e^+ ( p_2, \sigma_2) \rightarrow Z( k, \lambda ) + H( q ) ,
\end{equation}
where $\sigma_{1,2} = +\frac{1}{2}, -\frac{1}{2} $ and $\lambda = -1, 0, +1 $ .
The invariant amplitude for this Higgs production is
 \begin{equation}
 {\cal M}^\lambda = {\bar v}(p_1) ( v_e I + a_e \gamma_5 ) \gamma_{\tau} u( p_2 ) P^{\tau\mu} V_{\mu\nu} ( k + q , k ) \epsilon^{\lambda,\nu} ,
 \end{equation}
 where $P^{\tau\mu} $ is the propagator of virtual Z boson in unitary gauge  and  the polarization vector $\epsilon^\lambda (k) $ of real  Z   is
  \begin{eqnarray}
 &&\epsilon^{\pm,\mu}=(0, cos\hat \theta, \mp I, -sin\hat\theta)/\sqrt 2,\nonumber \\
  &&\epsilon^{0,\mu}=(k, E_z sin\hat\theta, 0, E_z cos\hat\theta)/M_z ,
   \end{eqnarray}
with $E_Z$   the energy of $Z$ boson.

 In the rest frame of real  Z,  the decay (helicity) amplitude is written as \cite{Hagiwara:1993sw}
 \begin{equation}
 D_{\lambda,\tau} ( k^2 , \vartheta, \varphi )  =  \sqrt {k^2 } ( v_f + \tau a_f ) d^\tau_\lambda ( \vartheta, \varphi) ,
 \end{equation}
 where $\tau$ is the helicity of the spin analyzer in Z decay, and $ d^\tau_\lambda ( \vartheta, \varphi)$ is the usual $\frac{1}{2}-$representation of rotation group.
 There is also  a Breit-Wigner form but left out as an overall factor.  The scatting angle ${\hat \vartheta}$, polarization angle $ \vartheta$ and azimuthal angle $\varphi$ is defined in   Fig.\ref{anglesdefi}. 

\subsection{ Total cross section for Higgsstrahlung }

 The differential cross section  for Higss production at Born aproximation reads
 \begin{eqnarray}
 \begin{aligned}
 \frac{ d \sigma }{ d cos {\hat \vartheta}  d cos\vartheta d\varphi } &= K  \sum_\tau   D^{*}_{\lambda ' ,\tau}  \rho^{ \lambda ' \lambda } D_{\lambda,\tau} \\
 &=  K  \sum_\tau {\bar  {\sum_{\sigma_1, \sigma_2} } }    D^{*}_{\lambda ' ,\tau}  { {\cal M}^{\lambda '}}^\dag  {\cal M}^\lambda D_{\lambda,\tau}  .
  \end{aligned}
 \end{eqnarray}
The       kinetic factor $  K$   reads :
\begin{equation}
K  =  \frac{ \beta ( m^2_Z/s, m^2_H/s  ) }{ 128   s  \left|     s- m^2_Z   \right| ^2   }
 \frac{ | \mathbf{p}_l  |   }{ 32 \pi^3 M_Z^2  \Gamma_Z} ,
\end{equation}
where  $\beta ( a, b  ) = {( 1 + a^2 + b^2 - 2a - 2b - 2 a b ) }^{1/2}$, with $s$   the center of mass energy square  and $| \mathbf{p}_l | $   the momentum of lepton as Z spin analyser.


After integration of phase space, the total cross section is :
\begin{eqnarray}
 \sigma 
              &= & K \frac{ 128  \pi C_{l{\bar l}} ~s~  }{ 9  } Q ,
 \end{eqnarray}
where
  \begin{equation}
 Q=   ( g^2_0 + 2   g'_3{g_0}  ) (E^2_Z  + 2 m^2_Z) +   \frac{1}{2} g_1 g_0   \beta^2 E_Z  s^{3/2}.\label{Q}
   \end{equation}
Since the  New Physics couplings are a small perturbation from the SM couplings, we keep only  the leading order linear term contributions.
It's interesting that, the  anomalous  couplings   appear as a combination:
\begin{equation}
g '_3 = 2 g_2 ( s + m^2_Z ) + g_3  + g_1 \sqrt{s} E_Z.
\label{newcoupling}
\end{equation}
This further reduces the number of free parameters to three, $g_1$, $ g '_3$ and ${\tilde g}$. We'd also like to point out that, this combination take place at the level of amplitude of ZH associating production, so it is regarded as a new parameterization for the analyzing of Higgsstrahlung channel. To isolate the $g_2$ contribution one has to  investigate the channel of Higgs's decaying to Z pair whose yields seems smaller, as become an independent issue beyond the scope of this work.

\subsection{ Polarization in $Z$ boson decay }

Although only three effective couplings left, one can not distinguish their contributions by only the total cross section measurement. Different kinds of new physics structure will give more information in the angular distributions of the decay products of Z boson, which characterize the polarization fractions of the Z boson.
The  polar angle distribution    of the outgoing lepton is derived as
 \begin{eqnarray}
  \frac{ d \sigma }{ \sigma  d \cos\vartheta }  & = &  \frac{3 M_Z^2 }{ 8  ( a^2_f+ v^2_f )~Q }\times      \nonumber    \\
 & ~& \left\{ \left[    \left(  g^2_0 +  2  { g '_3}{g_0 }  \right)  \frac{E_Z^2}{M_Z^2}  + g_1 g_0  \frac{Q_1}{M_Z^2}  \right]  \Gamma^0   (\vartheta)+ \right. \nonumber  \\
 &~&  \left.  \left(g^2_0+ 2 {g '_3}{g_0} \right)  [   \Gamma^-  (\vartheta) +  \Gamma^+  (\vartheta)    ]        \right     \} ,
 \label{polar}
\end{eqnarray}
where $\Gamma^\lambda  (\vartheta)  $ is the normalized partial width of Z boson in $\lambda$  helicity state,    defined as
\begin{eqnarray}
\Gamma^\pm  (\vartheta)  &=& 
 \frac{1}{2} M_Z^2 \left[ (a_f^2+v_f^2) ( \cos 2\vartheta +3 )  + 8~\pm~  a_f v_f \cos\vartheta  \right] \\
 \Gamma^0  (\vartheta)  &=&  2 M_Z^2 (a_f^2+vf^2) \sin^2\vartheta .  
 \end{eqnarray}
 The fraction  of each spin polarization characterized by the distribution of the polarization angle $\vartheta$, are obtained
by integrating out the scatting angle ${\hat \vartheta}$.  It is interesting to note that, the fraction of transverse polarization can be increased, if integration  of the scattering angle not in the entire region, for example,
a forward region can be defined by $ | cos {\hat \vartheta} | > cos \frac{\pi}{4} $,
 \begin{eqnarray}
  \frac{ d \sigma }{ \sigma  d cos\vartheta } |_{fwd} &=&\frac{3 M_Z^2 }{ 128  ( a^2_f+ v^2_f )~Q } \times \{ 2 ( 8-5\sqrt{2} )\cdot \nonumber \\
  && \left [     ( g^2_0 +  2  { g '_3}{g_0 }  )  \frac{E_Z^2}{M_Z^2}  + g_1 g_0  \frac{Q_1}{M_Z^2} \right ]  \Gamma^0   (\vartheta)  + \nonumber \\
   &&  ( 16 - 7\sqrt{2} )  (g^2_0+ 2  {g '_3}{g_0} )\times \nonumber \\
   && [   \Gamma^-  (\vartheta) +  \Gamma^+  (\vartheta)    ]  \}.
   \label{polarfwd}
\end{eqnarray}
It's obvious that the contribution from $ \Gamma^\pm  (\vartheta) $  is enhance by a factor of $
3.3$ in the forward region. 
 In experiments, this polarization distribution, together with the total cross section, will be used to fit the parameters  $g_1$ and $ g '_3$.

\subsection{ Azimuthal angle distribution for CP violation }

Up to now, all the   analyses are independent of the CP violation term $\widetilde{g}$ in the effective Hamiltonian of eq.(\ref{effH}). This term in the effective Hamiltonian characterizes the  CP violating effects in the new physics beyond the standard model. We have to study the azimuthal angle  $\varphi$ dependence in the Z boson decay in order to study this CP violation effects:
 \begin{eqnarray}
 \frac{ d \sigma }{ \sigma  d \varphi} &=&   \frac{ 1}{2 \pi  }  -
  \frac{ M_Z^2}{4 \pi Q}  \times \left\{   (  g^2_0 + 2  {g '_3}{g_0} ) \cos{2\varphi} 
  +g_0 {\tilde g} s \beta  \sin{2\varphi} 
  \right\}.
\label{eqazimuthal}
 \end{eqnarray}
Here
the first two terms came as a background from Standard Model,  CP-violation shows
up in the third term, whose $\sin2\varphi$ dependence signal itself against the background in the shape of $\cos 2\varphi$.

\begin{figure}
 \begin{center}
\includegraphics[width=8cm]{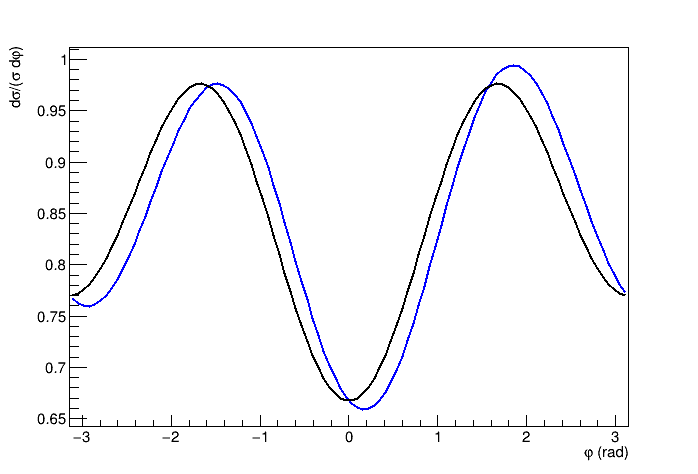}
\caption{\label{azimuthalfig}  Differential cross section of Higgs production as function of  azimuthal angle $\varphi$ by Eq.(\ref{fwdazimuthal}), black for SM while blue for possible new physics beyond SM }
\end{center}
\end{figure}

There is no  $\sin\varphi$ term in the above equation. However it can be recovered by breaking the symmetry in decay angle $\vartheta$ integration only  $0\to \pi/2$ or $\pi/2 \to \pi$,
at a price of $\cos\varphi$ background in the SM
 \begin{eqnarray}
  \frac{ d \sigma }{ \sigma  d \varphi} |_{\vartheta \gtrless \pi/2 } &=&   \frac{ M_Z^2}{ 16 \pi Q} \times  \left\{  \frac{8Q}{M^2_Z} - 4   ( g^2_0 + 2  {g '_3}{g_0} ) \cos{2\varphi} \right. \nonumber  \\
&& -4 g_0 {\tilde g} s \beta  \sin{2\varphi} \pm 3  g_0 {\tilde g} s \frac{\pi a_e v_e }{ v^2_e + a^2_e }\frac{E_Z}{M_Z}   \sin\varphi  \nonumber \\
 && \pm 6\frac{\pi a_e v_e}{ (v^2_e + a^2_e )} \left[  ( g^2_0  - 2 {g '_3}{g_0} ) \frac{E_Z}{M_Z}  + 
   \left. g_0 g '_3 \frac{Q_1}{ E_Z M_Z } \right]   \cos{\varphi}\right\}.
\label{fwdazimuthal}
 \end{eqnarray}
One can   see from Fig.\ref{azimuthalfig}, that this distribution with $\sin\varphi$ will signature CP violation by breaking the height-equality of two peaks in the background; while the $\sin 2\varphi$ term makes a phase shift against CP conserving backgrounds of the standard model.

\section{  Estimations of constrain limits at future Higgs  factories  }

 At future Higgs factories,     millions of Higgs production  events are expected, which will give signal of new physics or provide   at least   constrains to the new physics presented in the form of Eq.~(\ref{Feynrule}).
  For example, the circular electron position collider may deliver a luminosity of  $5000 {fb}^{-1}$ at center of mass energy $E=240 GeV$.  In the conceptual design report \cite{CEPCStudyGroup:2018ghi}, the exclusive channel of $e^-e^+ \to ZH \rightarrow l^+ l^- b {\bar b}$ is investigated with phase space cuts :
  \begin{itemize}
  \item  $p_l \ge 18GeV$, $p_b \ge 20 GeV$,
  \item $  | cos\theta_l | \le 0.98 , | cos\theta_b | < 0.98 $,
  \item $ | M_{l^+ l^-} - M_Z | < 15 GeV , | M_{b{\bar b}} - M_H  | < 12 GeV $ .
  \end{itemize}
 Furthermore the CEPC simulations provided the expected performance to use :
  \begin{itemize}
  \item  lepton identification efficiency :  $85\%$
  \item  bottom jet tagging efficiency : $75\%$ .
  \end{itemize}

We have adopted relatively tighter cuts on phase space and on particle tagging (identification) so that the background (mainly $ZZ$ production) can be suppressed to ignorable level,
at least their contamination can be well estimated and subtracted in future experiment.

Before the real Higgs factory data and the details of possible systematical studies become available, we simply do  the simulations with Monte Carlo comparing the new physics contribution with  that from standard model. Based on the histograms for the angle distributions in SM, a Pearson $\chi^2$ is defined simply with the events numbers by  hypothesis of new physics from each bin of angle histogram. When the parameters $g_1$, $g'_3$ and $ {\tilde g} $  reach to sufficiently small  magnitudes, the effect of new physics will be concealed beneath the coverage of (mainly statistical) SM errors (reflected by $\chi^2$) in the future experiment, so their limits of sensitivity can be estimated accordingly.

Instead of inviting additional assumptions or more complicated procedures, anywhere we quoted as well a systematical error at the same level as the statistical one, then the sensitivities limits estimated in following subsections are very conservative. Any optimized and reliable limit-setting should be left to the actual data analysis in future upon more reasonable experimental inputs.

\subsection{ Limit from total cross section }

\begin{figure}
\begin{center}
\includegraphics[width=9cm]{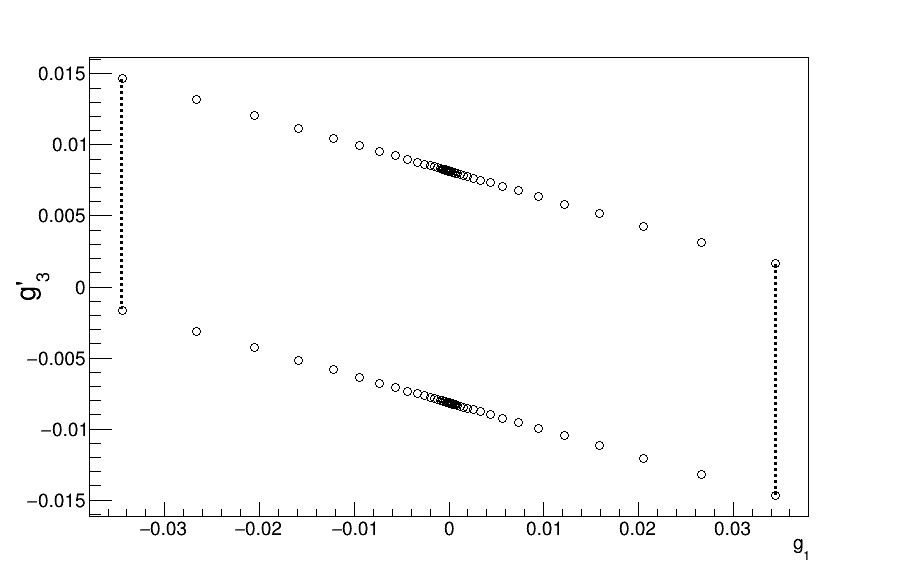}
\caption{\label{xseclim}  New physics sensitivity limits from total cross section measurements. The parameters     inside the parallelogram region are difficult to distinguish   from the standard model within experimental errors. }
\end{center}\end{figure}

Using the above mentioned cuts in the Higgs factory, we  scanned   the new physics parameters $( g_1 , g'_3 )$   simultaneously.   Their sensitivity limits will arrive when $\Delta \sigma / \sigma \ge \sqrt{2}/ \sqrt{N_{evt}} $, where   $ N_{evt}$ is the observed (signal) event number in  $ZH \rightarrow l^+ l^- b {\bar b}$, as shown in
  Fig.\ref{xseclim}.  The new physics parameters of $( g_1 , g'_3 )$    inside the parallelogram  will be difficult to distinguish   from the standard model within experimental errors. It can be understood that CEPC can set a limit down to $ | g'_3|  \le 0.015 $ and $ |g_1| \le 0.035$.

One may also set the limits lower to $ | g'_3|  \le 0.005 $ and $ |g_1| \le 0.015$ to pursue a higher sensitivity, by discarding the systematical errors and relaxing events selections. 

Just for the  total cross section,  on the other hand, the reconstruction of recoiled Z boson will lead to an inclusive analysis with Higgs decaying to anything rather than merely $ b {\bar b}$ final states. In this case, a tighter limit can be set with about 3 times larger statistics. In such a reconstruction  of recoiled Higgs, it's possible to walk around the dependence on the invisible decays from Higgs, however its details go beyond the scope of this paper.

\subsection{ Limit from Z polarization }

\begin{figure}[!htbp]
\begin{center}
\includegraphics[width=8cm]{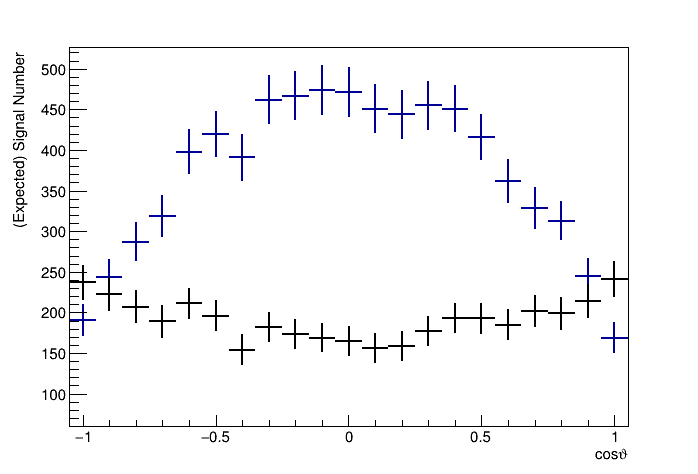}
\caption{\label{polangle}  Expected event number distributed along the polarization angle $\vartheta$, blue for Eq.(\ref{polar}) and black for Eq.(\ref{polarfwd}) }
\end{center}\end{figure}

When there is enough experimental data, we can also study the new physics effect sensitivity through polarization angle distribution shown in Eq.(\ref{polar}) and   (\ref{polarfwd}). The expected event number distributed along the polarization angle $\vartheta$,  are shown  in Fig.\ref{polangle}, with blue points for Eq.(\ref{polar}) and black points for Eq.(\ref{polarfwd}).
The polarization angle will distribute differently as the black plot in Fig.\ref{polangle} if only the forward region of the decay angle is investigated as in Eq.(\ref{polarfwd}).  Since it will come with lower statistics (only half number of   events), it will be skipped in the current numerical analysis, until there is   better input for experimental systematics.

\begin{figure}[!htbp]
\begin{center}
\includegraphics[width=9cm]{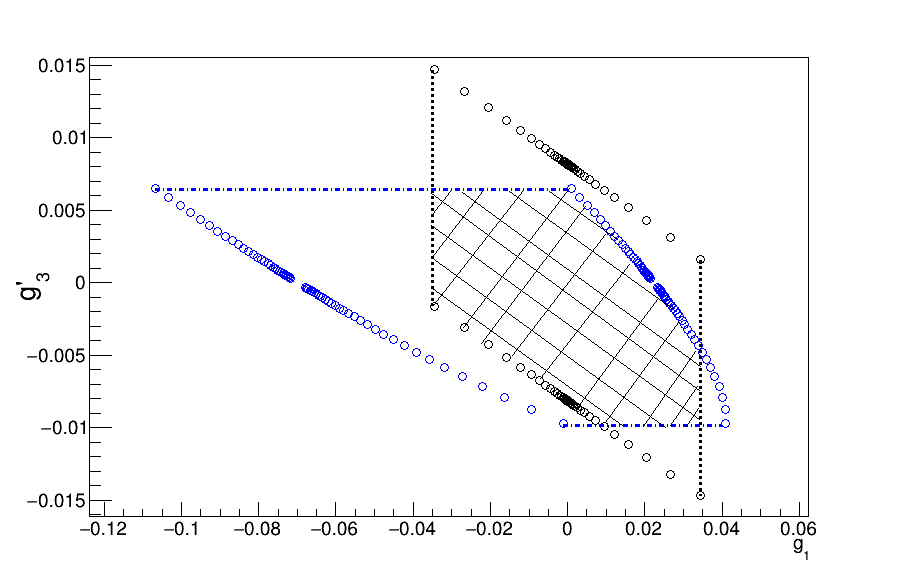}
\caption{\label{polalim}  Limits from polarization angle, no sensitivity in the belt closed by the blue lines. The black belt is from Fig.\ref{xseclim}. The overlap of two belts are in the meshed region }
\end{center}\end{figure}

After sensitivity study, we show the experimental limits from polarization angle distributions for the new physics parameters  $g_1$ and $g'_3$  in Fig.\ref{polalim}. Parameter regions inside  the blue lines are not distinguishable from the standard model.
We also copy the limits from cross section study in Fig.\ref{xseclim}. It  shows in Fig.\ref{polalim}  that,   the two limit regions have some overlaps and also some differences. This means that the sensitivity limits are further narrowed into the meshed region.
The polarization angle distribution will be anyway helpful since it will  constrain new physics from a different direction rather than the cross section. It's also worth to point out that, the distribution of polarization angle is   normalized as  in Eq.(\ref{polar}) by the   cross section.  It means less dependence or uncertainties from Higgs production or decay, because of the  actual analysis  by a fit solely on the shape. This indicates better determination of the HZZ couplings.

\subsection{ Limit for CP Violation parameter  $\tilde g$ }

\begin{figure}[!htbp]
\begin{center}
\includegraphics[width=8cm]{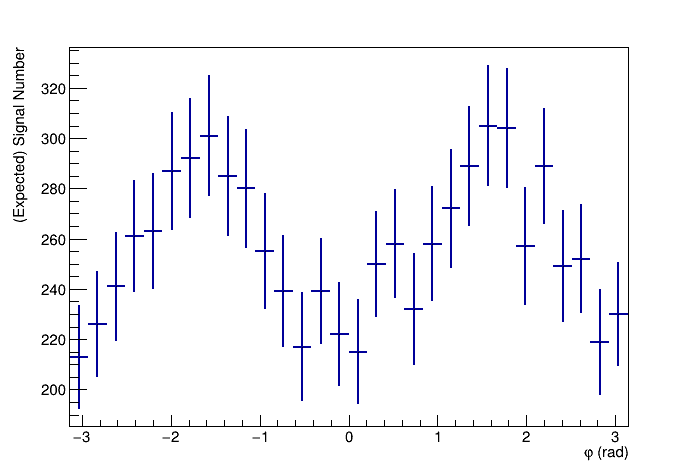}
\caption{\label{figazimuthal}  Expected event number distributed along azimuthal angle $\varphi$}
\end{center}\end{figure}

According to Eq.(\ref{eqazimuthal}), we show the expected event number distributed along azimuthal angle $\varphi$ in  Fig.\ref{figazimuthal}, where CP violation effect may show up.
Without losing of generality, each time only one of new physics parameters $g_1$ or  $g'_3 $ will be scanned together with the CP violation parameter $\tilde g$.

\begin{figure}[!htbp]
\begin{center}
\includegraphics[width=9cm]{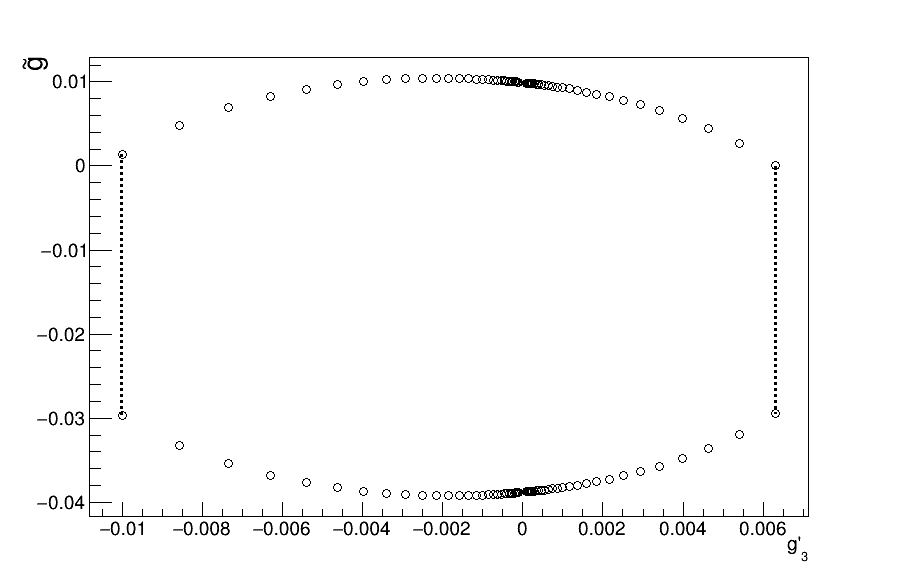}
\caption{\label{cpvg3}  Experimental limits of $\tilde g$ from azimuthal angle distribution study, with $g'_3$ in correlation, with no sensitivity in the belt }
\end{center}\end{figure}

\begin{figure}[!htbp]
\begin{center}
\includegraphics[width=9cm]{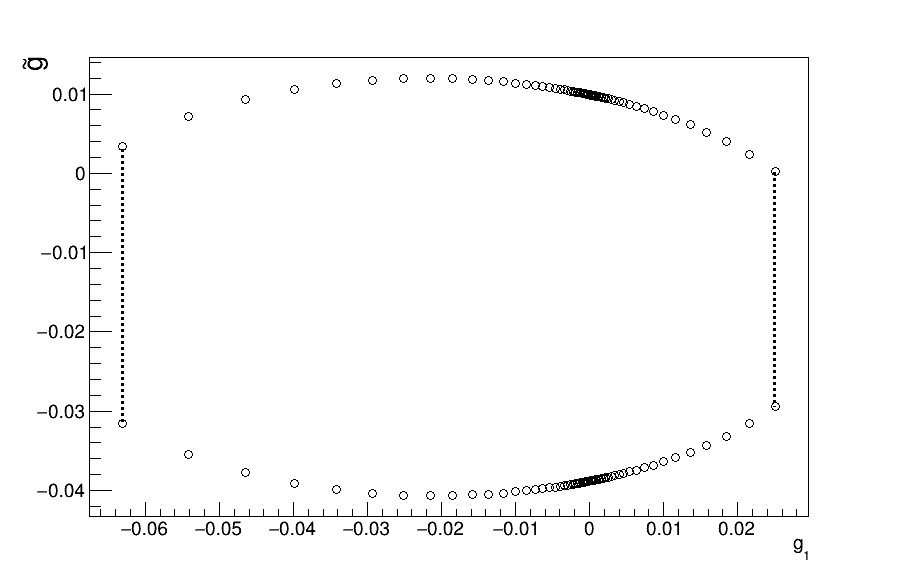}
\caption{\label{cpvg1} Experimental limits  of $\tilde g$ from azimuthal angle distribution study, with $g_1$ in correlation, no sensitivity in the belt }
\end{center}\end{figure}

Again the forward region defined in Eq.(\ref{fwdazimuthal}) will be skipped in the current study for its lower statistics. After careful study of  the backgrounds, we derive the experimental limit of $\tilde g$ with the correlation of $g'_3$ shown in   Fig.\ref{cpvg3}. The correlation sensitivity  of $\tilde g$  and $g_1$ are shown in Fig. \ref{cpvg1}. 
These figures indicate that, the experimental sensitivity can reach the limit of $\tilde g$ to $-0.04 \sim 0.01$.

\section{  Summary  }

We have studied the new physics sensitivity in the $e^+ e^- \rightarrow HZ $ process of the future Higgs factory. By the cross section and angular distribution measurements, we set experimental limits for the  Dimension-6 operators of  Effective Field Theory in a model independent way.  Especially by the study of azimuthal angle  distribution of the Z boson decay, we found that the future Higgs factory can  set a stringent limit to the   CP-violation effective operators in the new physics,   i.e. the $ {\tilde g}$  sensitivity limit up to $-0.04 \sim 0.01$. 
Our study shows that  the future electron positron collider    will be   an ideal machine for the search of New Physics signal.

\acknowledgments{We thank Prof. CP-Yuan and Dr. Yan Bin for useful discussions, we would like to thank Prof. Gong Bin for illuminating comments.
This work was supported in part by National
Natural Science Foundation of China under Grant No. 11521505 and 11621131001.
}







\vspace{-1mm}
\centerline{\rule{80mm}{0.1pt}}
\vspace{2mm}


\bibliographystyle{ieeetr}
\bibliography{Higgszz}

\clearpage

\end{document}